\documentclass[]{emulateapj}
\usepackage{graphicx}
\shorttitle{Density scalings in supersonic MHD}
\begin{document}
\title{Density scaling and anisotropy
 in supersonic MHD turbulence}            
\author{A. Beresnyak, A. Lazarian}
\affil{Dept. of Astronomy, Univ. of Wisconsin, Madison, WI 53706}
\email{andrey, lazarian@astro.wisc.edu}
\and
\author{J. Cho}
\affil{Chungnam National Univ., Korea}
\email{cho@canopus.chungnam.ac.kr}

\begin{abstract} 
We study the statistics of density for supersonic turbulence in a medium
with magnetic pressure larger than the gaseous pressure. This study is
motivated by molecular cloud research.
Our simulations exhibit clumpy density structures, which 
contrast increases with the Mach number.
At 10 Machs densities of some clumps are three orders of magnitude higher than
the mean density. These clumps give rise to flat and approximately isotropic density spectrum
corresponding to the random distribution of clumps in space. 
We claim  that
the clumps originate
from our random, isotropic turbulence driving. 
When the contribution from those clumps is suppressed by studying logarithm of density,
the density statistics exhibit scale-dependent anisotropy consistent
with the models where density structures arise from shearing by Alfv\'en waves.
It is noteworthy that originally such models were advocated for the case
of low-Mach, nearly incompressible turbulence. 

\end{abstract}


\section{Introduction}
The paradigm of interstellar medium has undergone substantial changes 
recently. Instead of quiescent medium with hanging and slowly evolving
clouds a turbulent picture emerged (see review by V\'azquez-Semadeni et al., 2000).
With magnetic field being dynamically important and dominating the gas
pressure in molecular clouds, this calls for studies of compressible
magnetohydrodynamic (MHD) turbulence.

Recent years have been also marked by a substantial progress in
understanding of the MHD turbulence statistics (see review
by Cho \& Lazarian 2005 and references therein). This statistics
allows to find regularities in turbulence, e.g. power spectrum
allows to learn how much energy is at a particular range of scales.

A very important insight into the
incompressible MHD turbulence by Goldreich \& Shridhar (1995) (henceforth GS95)
has been followed by progress in understanding of compressible MHD turbulence
(Lithwick \& Goldreich 2001, Cho \& Lazarian 2003, henceforth CL03, Cho, Lazarian \& Vishniac 2003, Vestuto et al. 2003).
In particular, simulation in Cho \& Lazarian (2003) showed that Alfv\'enic cascade evolves
on its own\footnote{
The expression proposed and tested in CL03 shows that the coupling of Alfv\'enic and
compressible modes is appreciable at the injection scale if the injection velocity
is comparable with the {\it total} Mach number of the turbulence, i.e. with
$(V_A^2+C_S^2)^{1/2}$, where $V_A$ and $C_S$ are the Alfv\'en and sound velocities
respectively. However, the coupling gets marginal at smaller scales  as
turbulence cascades and
turbulent velocities get smaller.}
and it exhibits Kolmogorov type scaling (i.e. $E\sim k^{-5/3}$)
and scale-dependent
anisotropy of the Goldreich-Shridhar type (i.e. $k_{\|}\sim k_{\perp}^{2/3}$)
even for high Mach number turbulence. While slow modes exhibit similar
scalings and anisotropy, fast modes show isotropy. The density scaling
obtained in Cho \& Lazarian (2003) was somewhat puzzling. At low Mach
numbers it was similar to slow modes, while it got isotropic 
for high Mach numbers.

The uncertainties associated with the earlier study motivate our present
one. Density statistics is important for understanding the structure
of molecular clouds and the associated processes of star formation.
Are density perturbations tend to be elongated along magnetic field
lines? How does this depend on Mach number of turbulence? These are the
questions that we would like to answer.

There are a number of observational  implications of the density
spectra.
Shallow power spectrum can result in a lot of small scale absorption 
(Deshpande 2000) which can account for the mysterious Tiny Scale
Structures or TSAS (Heiles 1997).
Power spectrum shallower than the Kolmogorov
one was reported in a number of observations (Deshpande, Dwarakanath \& Goss 2000, 
Padoan et al 2003). Can one explain this? Density anisotropies 
have been observed
in scintillation studies at small scales but it is unclear whether we should 
expect them at all scales.    

\section{Theoretical Considerations}
Subsonic compressible MHD is rather well studied topic today.
It is  suggestive that there may be an analogy between the subsonic
MHD turbulence and its incompressible counterpart, namely, GS95 model.
Therefore the correspondence between the the two revealed in CL03 is
expected.  

It could be easily seen, that in the low-beta case density is perturbed
mainly due to the slow mode (CL03). Slow modes are sheared by Alfv\'en turbulence,
therefore they exhibit
Kolmogorov scaling and GS95 anisotropy for low Mach numbers. However, for high
Mach numbers we expect shocks to develop. Density will be perturbed
mainly by those shocks. However, the relative perturbation of
density is likely to be proportional to density itself.

One can also approach the problem from the point of view of underlying hydrodynamic equations.
It is well known that there is a multiplicative symmetry with respect to density
in the ideal flow equations for an isothermal fluid (see e.g. Passot \& V\'azquez-Semadeni, 1998;
henceforth PV98). This assume that if there is some stochastic process disturbing the density
it should be a multiplicative process with respect to density, rather than
additive, and the distribution for density values should be lognormal,
rather that normal. 1D numerical simulations of high-Mach hydrodynamics confirmed
that the distribution is approximately lognormal, having power-law
tails in case of $\gamma\neq 1$ (PV98). 

In MHD, however, the above described symmetry is broken by the magnetic
tension. This could be qualitatively described as the higher density
regions having lower Alfv\'en speed, if we assume there is no significant
correlation between density and magnetic field. The latter is usually the case
with strongly magnetized, low beta fluid ($P_{\mbox{mag}}>P_{\mbox{gas}}$). It is interesting to
test whether this causes substantial deviation of distribution from log-normal law. 

In a low Mach turbulence the processes leading to the perturbation of density
are governed by the sound speed. Self evolution of those will be slow
in comparison with shearing by Alfv\'en waves.

With a high sonic Mach we expect a considerable amount of shocks arise.
In a sub-Alfv\'enic case, however, we expect oblique shocks be disrupted by
Alfv\'enic shearing, and, as most of the shocks are generated randomly
by driving, almost all of them will be sheared to smaller shocks.
The evolution of the weak shocks will be again governed by the sonic
speed, and structures from shearing as in low Mach case should arise.

We also note that shearing will not affect probability density
function (PDF) of the density, but have
to affect its spectra and structure function scaling. In other words,
we deal with two distinct physical processes, one of which, random
multiplication or division of density in presence of shocks, affect
PDF, while the other, Alfv\'enic shearing has to affect anisotropy
and scaling of the structure function of the density. 

In order to test this we performed direct numerical simulations.

\section{The code}
We used data cubes from our direct 3-dimensional numerical simulations
(see Cho \& Lazarian 2004). As we are interested in high-Mach turbulence,
we performed simulations on a periodic $512^3$ Cartesian grid
with the average sonic Mach numbers of $\sim 10$ and $\sim 3$.
The effects of numerical diffusion are expected to be important
at the scales of less than 10 grid points. 
We observed that, parallel to the magnetic field, velocities
stay supersonic down to 8 grid units for Mach 10, and 20 grid units for Mach 3.
We used the isothermal equation of state and randomly drove turbulence
on the scale about 2.5 times smaller than the box size.
The Alfv\'en velocity of the mean magnetic field was roughly the same as the rms velocity,
which correspond to Alfv\'enic Mach number of around unity.

\section{Results}
In Fig.~1 we show the distribution for log-density for various values
of Mach numbers for 3D numerical simulations of MHD. In all cases, except
subsonic $\beta$ was chosen so that Alfv\'enic Mach number $M_A$ is slightly
less then unity. This was motivated by the idea, that in a strongly
super-Alfv\'enic fluid, given enough time, magnetic field will grow,
approaching equipartition. As soon as we observe scales smaller than a driving
scale, we will see mildly super-Alfv\'enic or sub-Alfv\'enic turbulence.

\begin{figure}
\figurenum{1}
  \plotone{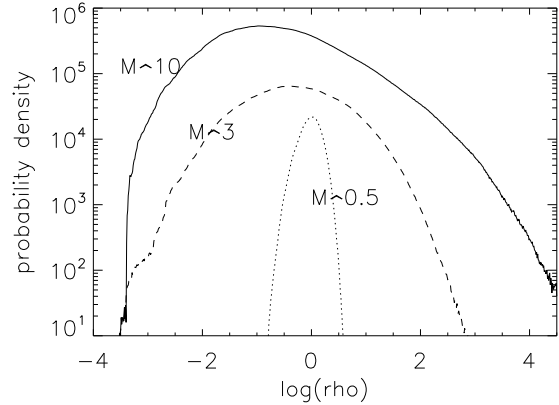}
  \caption{Probability density function for a density in direct
numerical simulation with Alfv\'enic Mach number around unity and various sonic
Mach numbers.}
\label{rhodist}
\end{figure}

\placefigure{rhodist}

We see, that the distribution shows significant deviation from a lognormal
law. The rms deviation of density for a subsonic case is consistent with
prediction $M^2$ for low beta (CL03), and the rms deviation of log-density
for supersonic case is around unity regardless of a Mach number. The distributions
are notably broader for higher Mach numbers, though. This is an indication that a
distribution is not universal.

Dimension of the high-density structures was between 1 and 2, being
viewed as a flatted filaments or elongated pancakes. There were no evident
preferred orientation of these structures along or perpendicular to
the mean magnetic field. Maximum density value in a Mach 10
data cube was around $3\times10^3\rho_0$.

It is obvious that density clumps with values of 3-4 orders of magnitude of mean density
can severely distort power spectrum. 
It is expected that these clumps can hide density structures created by motions at small
scales. We see in Fig.~2 that the power-law spectrum of density for high-Mach is very shallow.
Randomly distributed high-density clumps will also suppress
any anisotropy originating from motions at small scales.

\begin{figure}
\figurenum{2}
\plotone{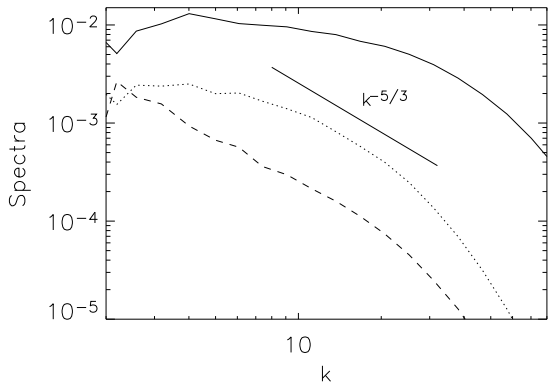}
 \caption{Mach number is 10, power spectra
 of: {\it solid} -- density, {\it dashed} -- velocity, {\it dotted}
-- logarithm of density.}
\label{spectrum}
\end{figure}

\placefigure{spectrum}

We attempted to overcome this effect and reveal an underlying density scalings by
using a log-density instead of density\footnote{density is dimensionless, normalized
by the mean density} for spectra and structure functions. 
We  found that this way to suppress
the influence of the high peaks to the  spectrum or structure function is superior to other
filtering procedures\footnote{Note that a nonlinear transformation can not change
function from isotropic to anisotropic and vice-versa. However anisotropy can be
substantially suppressed and not noticeable.}. Indeed, if we cut off peaks at some level,
it give similar results, but structure function looks worse, as the procedure of cutting
off, or restricting density to some level, introduces artificial structures in the real space.

\begin{figure*}
\figurenum{3}
  \includegraphics[width=0.32\textwidth]{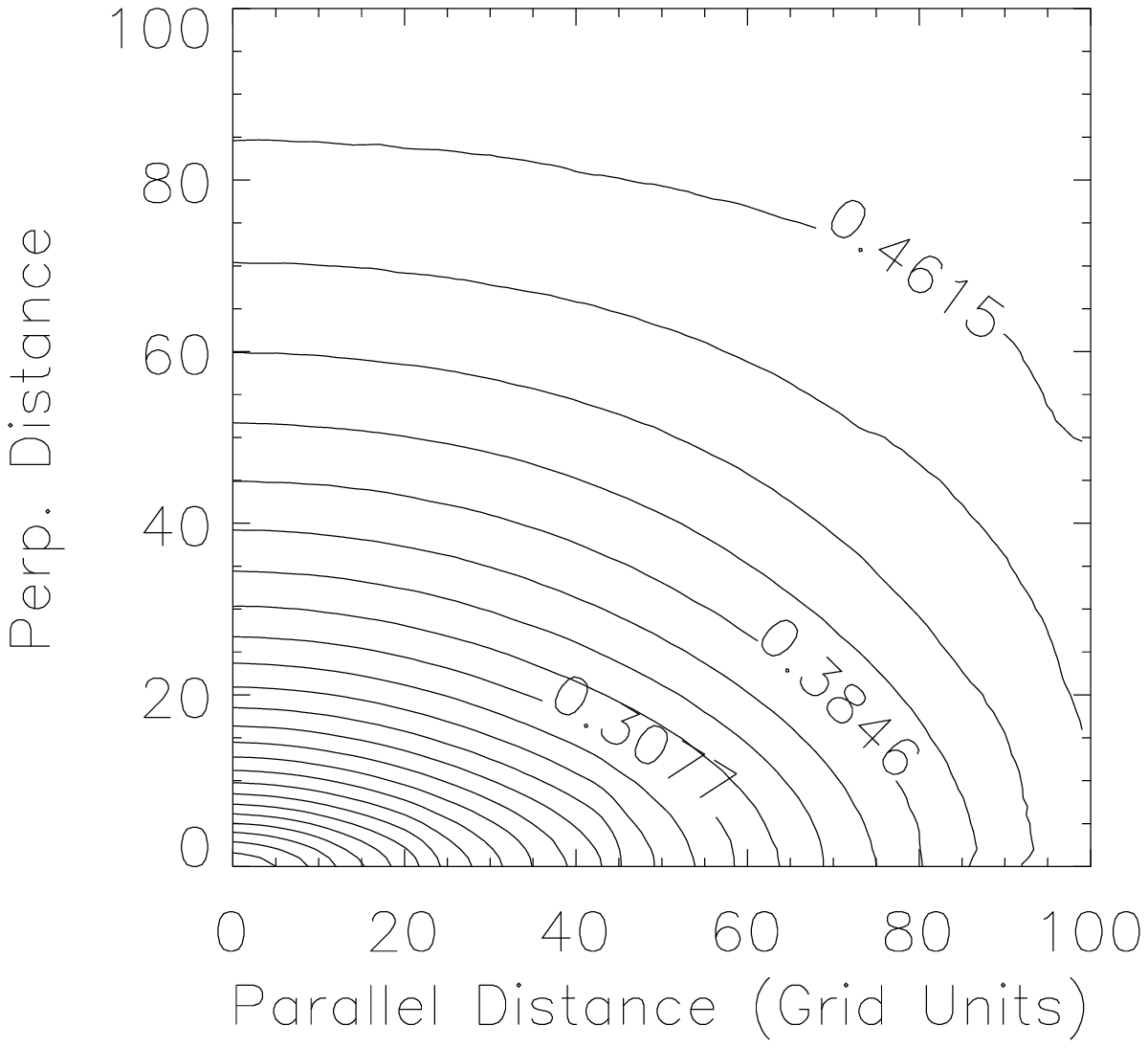}
  \hfill
  \includegraphics[width=0.32\textwidth]{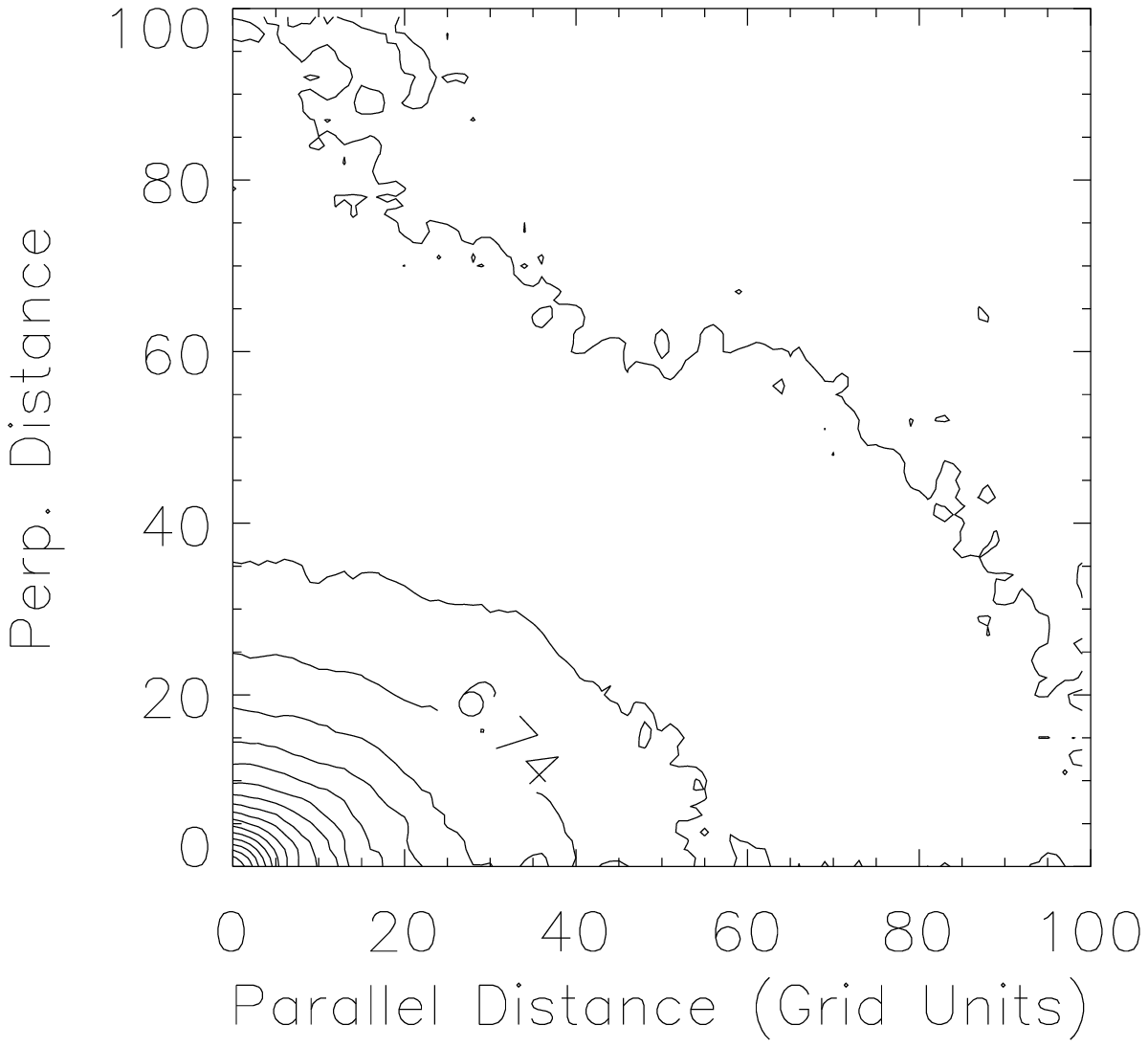}
  \hfill
  \includegraphics[width=0.32\textwidth]{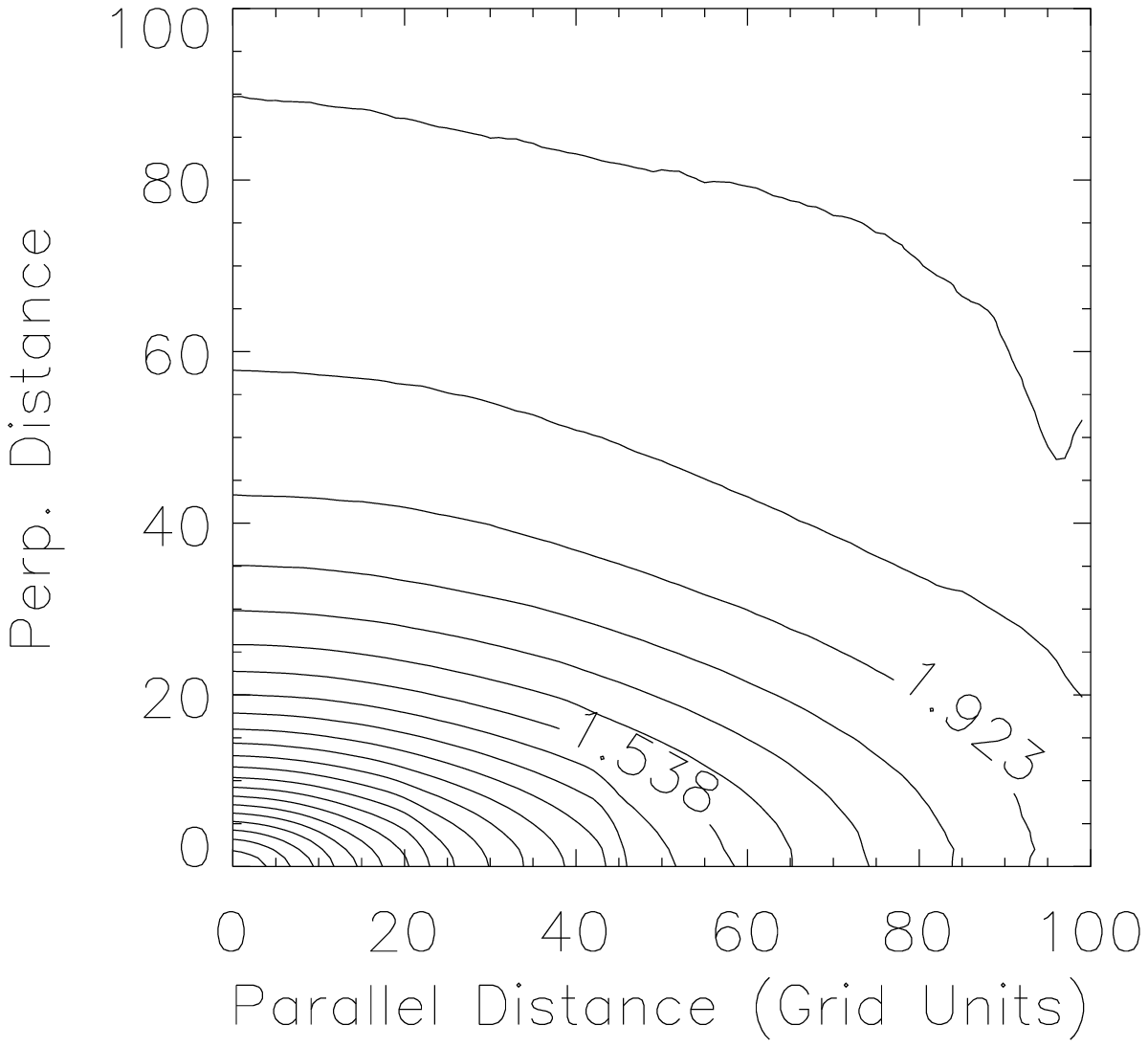}
  \caption{ $M_s\sim 10$, second order structure functions, calculated respective to
the {\it local} magnetic field for, {\it left panel}: magnetic field,
 {\it central panel}: density,
{\it right panel}: log-density. We see that in left and right panel structures are mostly elongated
parallel to the magnetic field. Structure function, e.g. for magnetic field,
is $|{\bf B}({\bf r}_0+{\bf r})-{\bf B}({\bf r}_0)|^2$, averaged over ${\bf r}_0$}
\label{SF}
\end{figure*}

\placefigure{SF}

The results and the comparison with the scalings of the magnetic field is presented in Fig.~3.
Magnetic field, being perturbed mostly by the Alfv\'en mode, shows GS95 anisotropy.
Scaling of log-density is analogous to it. We also see, that such a behavior is seen
right after the driving scale (which is around 100 grid units) by the magnetic field,
but somewhat into smaller scales by the log-density.

In Fig.~4 we checked for the scale-dependent anisotropy of the GS95 type 
($r_{\|}\sim r_{\perp}^{2/3}$). 

\begin{figure*}
\figurenum{4}
\plottwo{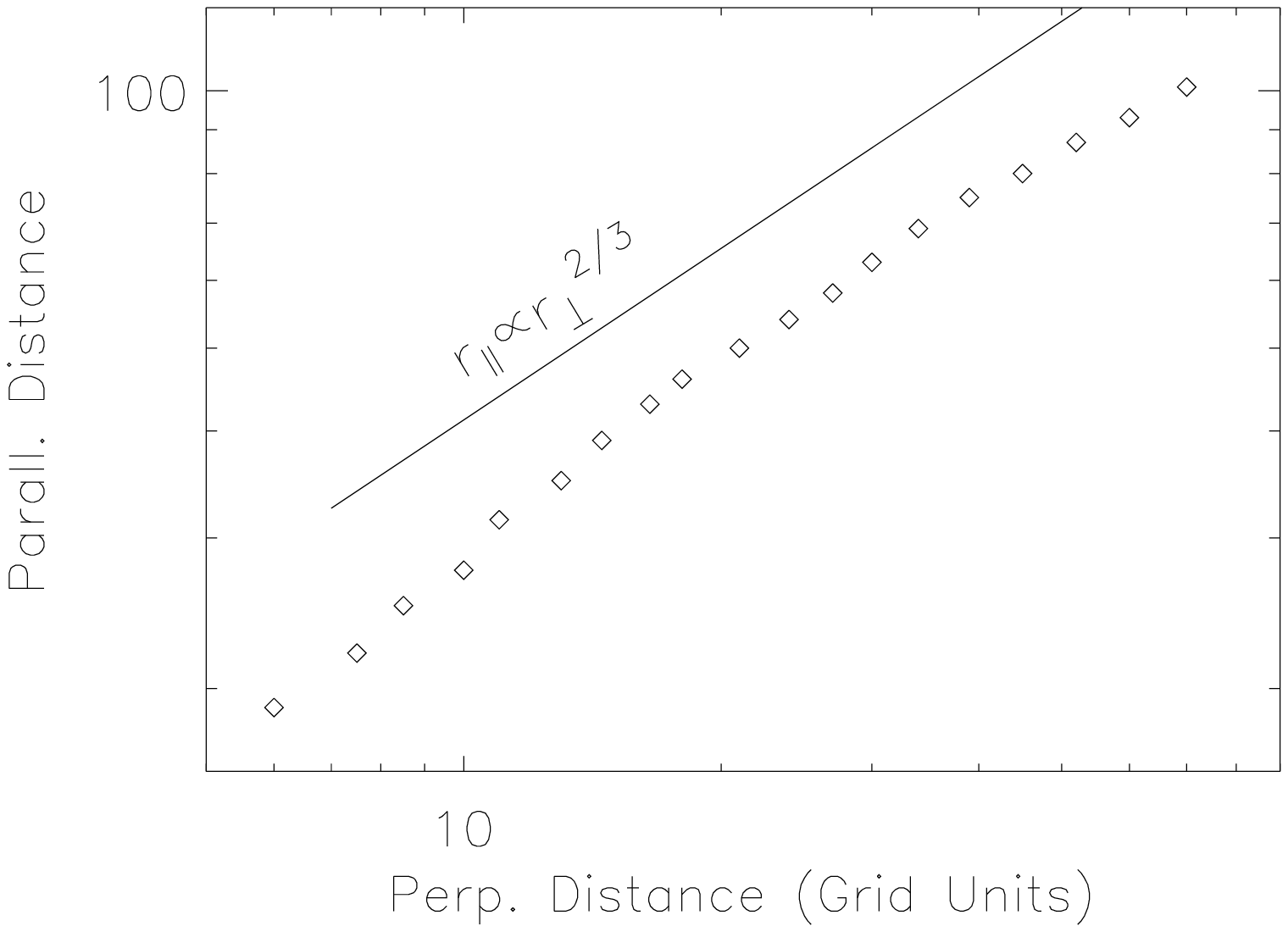}{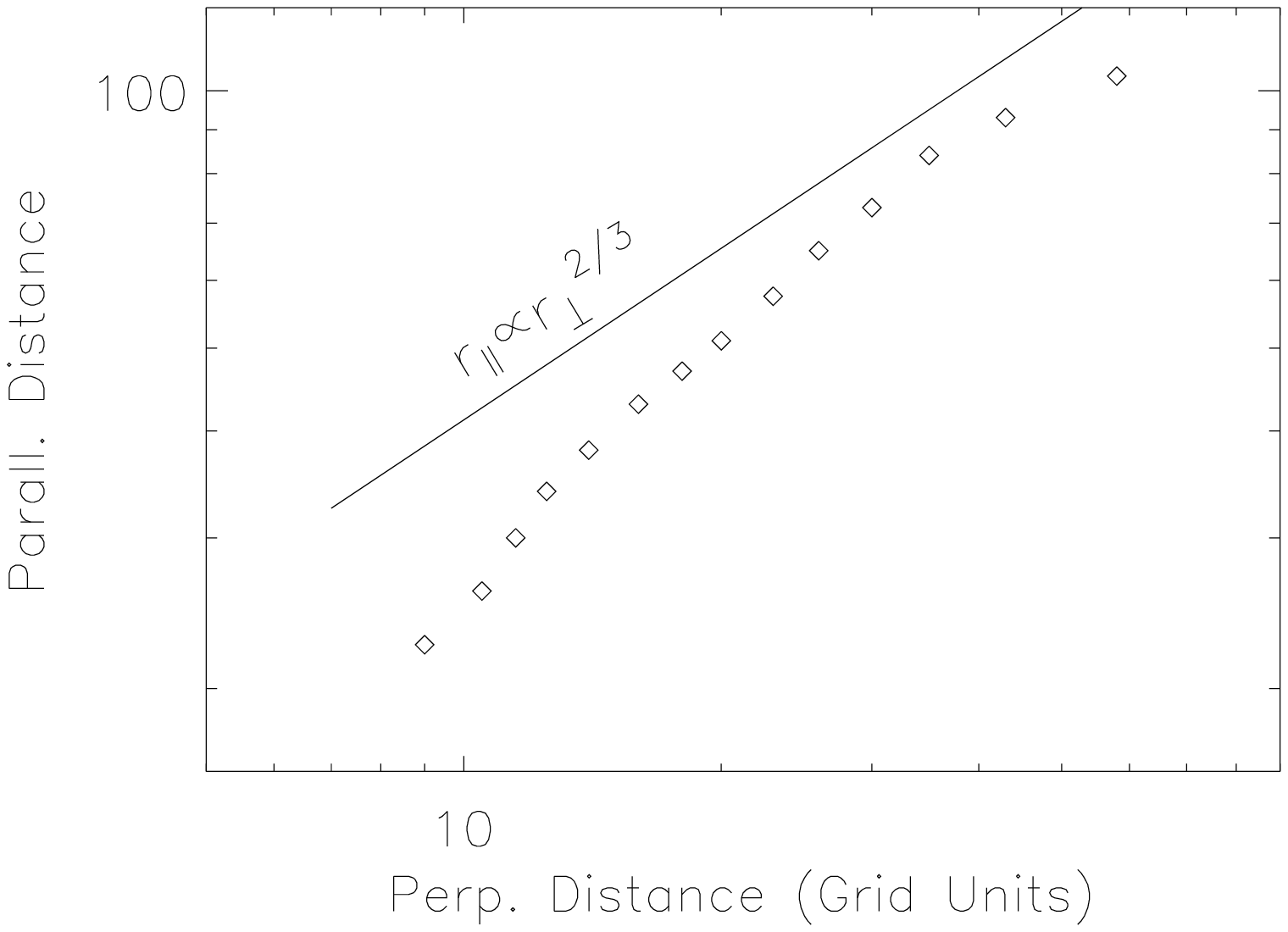}
 \caption{$M_s\sim 10$, values of $r_\|$ and $r_\perp$ with equal structure function
for {\it left panel}: magnetic field,
 {\it right panel}: log-density}
\label{anis}
\end{figure*}

\placefigure{anis}

We also checked for correlation between density and magnetic field
magnitude which was expected from a models of external compression of an
ideal MHD fluid, and in fact observed in many super-Alfv\'enic
simulations (see, e.g., Padoan \& Nordlund 1999). 
We have not found any significant correlation of this type.

\section{Discussion of Results}
It is a well known notion that a supersonic turbulence consists
mostly of shocks and other discontinuities. Our driving
is incompressible, but the modes are not decoupled at the
injection scale when Alfv\'enic Mach number is of the order of unity
(CL03). Therefore we expect that the driving excites an appreciable
amount of compressible motions. Indeed, 
our testing of data showed that
the rms velocity, associated
with slow mode was 
of the same order as the velocity of the Alfv\'en
mode. We assume that the resulting flat power spectrum of density
is associated with very large perturbation of density from compressible
motions that naturally arise at the driving scale due to coupling of compressible
and incompressible motions quantified in CL03.
Shocks in isothermal fluid can have very large density contrasts,
up to sonic Mach squared and can act as shocks in the snowplow phase
of supernova, namely, they collect matter keeping the total momentum
of the shock constant (see, e.g., Spitzer 1978). 
However, we do not see strong shocks near density clumps.
In magnetically dominated medium that we deal with it is reasonable
to assume that the corresponding shocks move material along magnetic
field lines the same way that the slow modes do in subsonic case.
The shocks are randomly oriented and therefore the clumpy structure
that we observe does not reveal any noticeable anisotropy.
Density perturbations associated with such shocks should not 
be correlated with the magnetic field strength enhancement similar
as in the case of densities induced by slow modes (see CL03).
Our analysis of the data confirms this. 

If we associate the clumps in simulations with interstellar clouds, in 
ISM with random driving we would expect the clouds not to be particularly
oriented in relation to magnetic fields at least until self-gravity does
get important. We observed substantial variation of the gas pressure, of three
orders of magnitude which is consistent with findings in Jenkins 2002. 
Flat spectrum observed is roughly consistent with some observational data.
Needless to say, that a more systematic analysis of data is required now
when we have theoretical expectations to test, e.g. the change of the
density spectrum with the Mach number. Testing the anisotropy of
density is another interesting project, even though one cannot directly
observe log-density and the effects associated
with the projection along the line of sight must be considered carefully
(see discussion of this in Esquivel et al. 2003).

Surely, for real clouds self-gravity can be important. This effect should
make the observed spectrum even flatter, as the density peaks will become
higher and more delta-function-like.
In addition, cooling may make
interstellar gas more pliable to compression than the isothermal gas that
we used in the simulations. 
This, could result in more density contrast when
the original gas is warm. 
However, usually interstellar warm gas has Mach
number of the order of unity. For molecular clouds for which Mach number
can be substantial our isothermal calculations seems to be adequate.

The alternative to this paper hypothesis explaining flat spectrum and isotopy,
namely that high-Mach large scale driving could affect turbulence to the
scales where effective Mach number is much less then unity is rather counter-intuitive
and contradicts the notion of the turbulence locality in $k$ space. 
We understood
that this unusual behavior is due to the fact, that density in high Mach simulations is
perturbed significantly nonlinearly, therefore making  
the interpretation of the power spectra more involved.
We used filtering to mitigate this effect,
and succeeded in showing that density scaling is anisotropic.
The range of scales where incompressible turbulent theory is applicable
is shortened in numerical simulation with supersonic driving. Between sub- and supersonic
scales there is a region where compressible motions cascade in a way that is 
yet to be
understood.

There exist a different effect that can make density spectrum flat at small scales.
Incompressible fluid with viscosity much larger than magnetic diffusivity, i.e.
in high Prandtl number fluid, at the viscosity-dominated scales, shows a peculiar
regime cascade reported in Cho, Lazarian \& Vishniac (2002).
If the the magnetic pressure associated with anisotropic magnetic filaments
formed in this new regime of MHD turbulence is balanced by the thermal pressure, 
it will create density structures with shallow $k^{-1}$ spectrum. 
However, this effect is present only at small scales where viscosity
by neutrals gets important. This allows to distinguish the two different causes of
spectrum flattening.

Most incompressible turbulence theories assume constant density,
everywhere. As we see, the supersonic driving provide transient, high contrast density
structures, as the back-reaction of the fluid is the term
$\nabla P/\rho$ which allow for very high density perturbations. This should not be
a problem for theories of strong turbulence where wavepackets are cascaded in a time
comparable with inverse frequency.
Weak turbulent theories, which have a cascading time $t_{\mbox{cas}}$
dependent on energy flow in k-space seem to be less applicable, as the Alfv\'en perturbation
has to travel $v_A t_{\mbox{cas}}$ distance to fully distorted. On the other hand,
$t_{\mbox{cas}}\propto\epsilon^{-1/2}$
for three-wave processes (Zakharov, L'vov \& Falkovich, 1992), where $\epsilon$ is the flux of energy in $k$ space.
This distance might be pretty long, depending on $\epsilon$ and exceed the lengthscale
at which density changes significantly. Therefore, ``normal'' weak turbulence will be pushed
on scales much smaller than subsonic scale. On the other hand, we don't expect weak turbulence
to be valid on smaller scales due to the increase of $k_\perp/k_\|$ with smaller scales
and the onset of strong Alfv\'enic turbulence.

\section{Acknowledgments}
AB thanks IceCube project for support of his research.
AL acknowledges the  NSF grant AST-0307869 and the support from
the Center for Magnetic Self-Organization in Laboratory and Astrophysical
Plasma.

\end{document}